\documentclass[showpacs,aps,amsmath,amssymb,twocolumn]{revtex4-2}
 
\usepackage[]{graphicx}
\usepackage{color}
\usepackage{bbold}
\usepackage{braket}
\usepackage[colorlinks=true, pdfstartview=FitV, linkcolor=blue, citecolor=red, urlcolor=black]{hyperref}
\newcommand{\be}{\begin{equation}}
\newcommand{\ee}{\end{equation}}
\newcommand{\ben}{\begin{eqnarray}}
\newcommand{\een}{\end{eqnarray}}
\newcommand{\bes}{\begin{subequations}}
\newcommand{\ees}{\end{subequations}}
\def\bal#1\eal{\begin{align}#1\end{align}}

\newcommand{\bfi}{\begin{figure}}
\newcommand{\efi}{\end{figure}}
\newcommand{\bc}{\begin{center}}
\newcommand{\ec}{\end{center}}
\newcommand{\sech}{\mbox{sech}}

\newcommand{\sn}{\mbox{sn}}
\newcommand{\dn}{\mbox{dn}}

\begin{document}
\title{Mechanism to control the internal structure of thick brane}
\author{D. Bazeia$^{1}$ and A. S. Lob\~ao, Jr.$^{2}$}
\affiliation{$^1$Departamento de F\'\i sica, Universidade Federal da Para\'\i ba, 58051-970 Jo\~ao Pessoa, PB, Brazil\\
$^2$Escola T\'ecnica de Sa\'ude de Cajazeiras, Universidade Federal de Campina Grande, 58900-000 Cajazeiras, PB, Brazil}
\begin{abstract}
In this work we study braneworlds generated by several scalar fields. The investigation describes the necessary formalism to examine models and evaluate the conditions for the stability of the gravitational sector under linear perturbations. In particular, we develop a mechanism that help us investigate distinct situations controlled by two and by three fields, focusing on how the fields can be used to modify the internal structure of the  brane. \end{abstract}

\maketitle
{\bf {\it{1. Introduction.}}}\, Braneworld models are in general five-dimensional theories of gravity theorized in the late 1990s as candidates to solve the hierarchy problem \cite{Randall:1999vf}. In this scenario the brane in thin, but it was soon changed to include scalar field to control the fifth dimension of infinite extent. The presence of the scalar field makes the brane thick, leading to a significant change in the brane structure \cite{Goldberger:1999uk,Skenderis:1999mm,Csaki:2000fc,DeWolfe:1999cp,Brito:2001hd}. Since the beginning of the year 2000, a lot of work on brane has appeared, adding specific contributions to the subject. An incomplete list of investigations can be found in \cite{A,B,Bb,C,D,Dd,E,Ee,F,G,H,Bazeia:2013uva,Bazeia:2015oqa}, in the reviews \cite{R1,R2,R3} and in other references therein.

In the thick braneworld scenario, a particularly interesting result is that the presence of the source scalar field which is controlled by the potential $V(\phi)$ may be described by the introduction of an auxiliary function $W=W(\phi)$, with the potential given by
\be\label{poten}
V=\frac12\, W_\phi^2-\frac43\, W^2,
\ee
where $W_\phi=dW/d\phi$. In this case, which is sometimes known as the first-order formalism, the equations of motion are solved via first-order differential equations. This formalism has appeared before in several works, for instance, in Refs. \cite{MC,Skenderis:1999mm,DeWolfe:1999cp}, and below we revisit the formal calculations in the case of several scalar fields, before advancing to describe the main results of the present work.  

In models described by several fields, both $V$ and $W$ in general depend on all the fields, and here we follow the suggestion considered in \cite{Bazeia:2006ef}, which takes the auxiliary function $W$ as the sum of distinct functions in the form $W=W_1(\phi_1)+ W_2(\phi_2)+\cdots+W_n(\phi_n)$. In the braneworld context which we are considering, the presence of the second term in the above potential preserves the first-order formalism, introduces interactions among the scalar fields and maintains the model nontrivial. This procedure was also investigated in Refs. \cite{Ahmed:2012nh,deSouzaDutra:2014ddw}, but we now go further and explore new features, not seen before. We focus, in particular, on the possibility of changing the internal structure of the brane. To do this, below we describe the general formalism and then consider several distinct models, controlled by the source scalar fields to induce modifications in the internal structure of the brane. 

{\it{2. Generalities.}}\, The thick braneworld models to be considered in this work are described by the action
\be\label{action}
{\cal S}=\int d^4 x dy \sqrt{|g|} \left(-\frac14\,R + {\cal L}_s\!\left(\phi_i,\nabla_a\phi_i\right)\right),
\ee
where the coordinate $y$ describes the extra dimension and $\phi_i\, ( i=1,2,\cdots,n) $ stands for the $n$ real scalar fields. Also, we use $4\pi G^{(5)}=1$ and $g=det(g_{ab})$ for $a,b=0,1,\ldots,4$, and the source Lagrange density has the standard form
\be
{\cal L}_s=\frac12\nabla_a\phi_i\nabla^a\phi_i-V,
\ee
where the potential $V$ in general describes the scalar fields $\phi_i$. Under standard investigation, the energy-stress tensor has the form
\be
T_{ab}=\nabla_a\phi_i\nabla_b\phi_i- g_{ab}\, {\cal L}_s,
\ee
and the equations of motion for the scalar fields are given by
\be
\nabla_a \nabla^a \phi_i+V_{\phi_i}=0,\;\;i=1,2,\cdots, n\,,
\ee
with $V_{\phi_i}=dV/d\phi_i$. Moreover, the Einstein equation is
\ben
R_{ab}-\frac12\, {R}\, g_{ab}=2\,T_{ab},
\een
where $R_{ab}$ is the Ricci tensor and $R$ the Ricci scalar.

In the braneworld scenario, the line element for the five-dimensional spacetime can be written as 
\be
ds^2_5=e^{2A}\,\eta_{\mu\nu}\,dx^\mu dx^\nu-dy^2,
\ee
where $\eta_{\mu\nu}\!=\!(1,\!-1,\!-1,\!-1)$ describes the four-dimensional Minkowski geometry, with $\mu,\nu$ running from $0$ to $3$, and the function $A$ controls the warp factor $\exp{(2A)}$.

A simple interpretation can be made when considering static configurations. In this case, we assume that the scalar fields and $A$ are static and only depend on the extra dimension; thus, $A=A(y)$ and $\phi_i=\phi_i(y)$ and the equations of motion for scalar fields reduce to
\be\label{fieldeqs}
\phi_i''+4A'\phi'_i=\,V_{\phi_i},\;\;i=1,2,\cdots, n,
\ee
where the prime denotes derivative with respect to extra dimension. Furthermore, the non vanishing components of the  Einstein equation become
\bes
\label{EinEq}
\ben
A^{\prime\prime}&=&-\frac{2}{3}\sum_i \phi_i'^2,\label{CompEins01}\\
A^{\prime2} &=& \frac{1}{6}\sum_i\phi_i'^2+\frac13V(\phi_i)\,.\label{CompEins02}
\een
\ees 
An important characteristic of the brane is its energy density, which is given by
\be\label{EnergyDensity}
\rho(y) = e^{2A(y)}\left(\frac12\,\sum_i \phi_i'^2+V\right)\,.
\ee 

We must now solve the set of second-order differential equations given by Eqs. \eqref{fieldeqs} and \eqref{EinEq} with suitable boundary conditions. This can be simplified by introducing the auxiliary function $W$ which in general depends on all the scalar fields to construct the first-order formalism: we suppose that
\be\label{FOE}
\phi_i^{\prime}=W_{\phi_i},\;\;i=1,2,\cdots, n,\quad\quad A^\prime=-\frac23 W\,,
\ee
which solve the Eqs. \eqref{fieldeqs} and \eqref{EinEq} when one writes the potential in the form
\be\label{eqPotV}
 V=\frac12 \sum_i W_{\phi_i}^2-\frac43\, W^2\,,
\ee
which is a generalization of Eq. \eqref{poten} to the case of several fields. Here the energy density becomes
\be\label{EnergyDensityFO}
\rho(y) = \frac{d}{dy}\left(e^{2A}W\right).
\ee 
To guaranty $4D$ gravity localization along the extra dimension, the warp factor has to vanish asymptotically. Thus, 
if $W$ does not diverge asymptotically, the location of the $4D$ gravity leads to a braneworld with vanishing energy, which is the scenario to be studied in this work.

Before dealing with specific models, let us first describe the conditions for stability of the general model. In this case, we follow \cite{DeWolfe:1999cp} and assume that the metric tensor is perturbed according to 
\be
ds^2=e^{2A}\left(\eta_{\mu\nu}+h_{\mu\nu}\left(x^\alpha,y\right)\right)dx^\mu dx^\nu-dy^2,
\ee 
where $h_{\mu\nu}\left(x^\alpha,y\right)$ depends on the extra dimension $y$ and on the four-dimensional position vector $x^\alpha$, which only acts on the four-dimensional part of the metric tensor.  Moreover, we assume that $h_{\mu\nu}$ satisfies the transverse and traceless conditions $\partial^{\mu} h_{\mu\nu}=0$ and $h_{\mu}{}^{\mu}\!=0$. We also consider fluctuations in the scalar fields as $\phi_i\to\phi_i(x)+\delta{\phi}_i\left(x^\alpha,y\right)$. Using these prescriptions in the equations of motion, the fluctuations in metric tensor decouple from the fluctuations in source fields and we get
\be\label{Eqstab01}
\left(\partial_y^2+4A'\partial_y\right)h_{\mu\nu}=e^{-2A}\Box^{(4)} h_{\mu\nu}\,.
\ee 
We now change coordinates from $y$ to $z$, so that $dz~=~e^{-A(y)}dy$ and rewrite $h_{\mu\nu}=e^{i \omega\cdot x }e^{-3A(z)/2} H_{\mu\nu}(z)$. With this, the Eq. \eqref{Eqstab01} can be represented by a Schrodinger-like equation 
\be\label{Schrodinger}
\left(-\frac{d^2}{dz^2}+{\cal U}(z)\right)H_{\mu\nu}=\omega^2 H_{\mu\nu}\,,
\ee 
where
\be
{\cal U}(z)=\frac32A_{zz}+\frac94A_{z}^2\,.
\ee 
In this representation we can write the Eq. \eqref{Schrodinger} as $S^{\dagger}SH_{\mu\nu}=\omega^2H_{\mu\nu}$, where $S=-d/dz-3A_z/2$. This shows that the braneworld scenario is stable against fluctuations in the metric, because Eq. \eqref{Schrodinger} cannot support negative energy bound states. We can also calculate the zero mode as
\be
H_{0}(z)=N\,e^{3A(z)/2}\,,
\ee 
where $N$ is the normalization constant. A transformation from $z\to y$, back to the variable $y$ leads to the stability potential in the form
\ben
{\cal U}(y)=\frac{3}4e^{2A}\left(2 A''+5 A'^2\right).\label{eqPotEsy}
\een

{\it{3. Specific Models.}}\, To see how the formalism presented here works, let us consider explicit examples of models described by two and by three real scalar fields.

\begin{center}{\it{Two fields}}\end{center}

The first model described by two real scalar fields $\phi$ and $\chi$ is defined by
\be
W(\phi,\chi)=r\!\left(\phi-\frac13\phi^3\right)+s\!\left(\chi-\frac13\chi^3\right),
\ee 
where $r$ and $s$ are real parameters which controls the influence of the fields on the braneworld. In the above braneworld scenario, the two fields are coupled since the term $W^2$ in Eq. \eqref{eqPotV} produces  interactions between them. Explicitly, the potential has the form
\be
\begin{aligned}
V(\phi,\chi)=&\,\frac12 r^2\!\left(1-\phi^2\right)^2+\frac12 s^2\!\left(1-\chi^2\right)^2\\
&-\frac43\!\left(r\!\left(\!\phi\!-\!\frac13\phi^3\!\right)\!+\!s\!\left(\!\chi\!-\!\frac13\chi^3\!\right)\!\right)^{\!2}\!.
\end{aligned}
\ee

We can use the first-order equations \eqref{FOE} to obtain the equations of motions of the scalar fields as
\bes\label{EPOMA}
\bal
\phi'&=r\left(1-\phi^2\right),\\
\chi'&=s\left(1-\chi^2\right).
\eal
\ees
A set of solutions that simultaneously obey the first-order equations \eqref{EPOMA} and the second-order equations \eqref{fieldeqs} are given by
\bes\label{soluMA}
\bal
\phi(y)&=\,\tanh\big(r(y-a)\big),\\
\chi(y)&=\,\tanh\big(s(y-b)\big),
\eal
\ees
where $a$ and $b$ are real constants which describe the centers of the kink-like solutions. Here we take $b=-a$ and study configurations that are symmetrically centered around the origin. We use this and the thickness of the scalar field solutions to show how modifications in these parameters may contribute to modify the profile of the brane.

\begin{figure}[t]
    \begin{center}
        \includegraphics[scale=0.6]{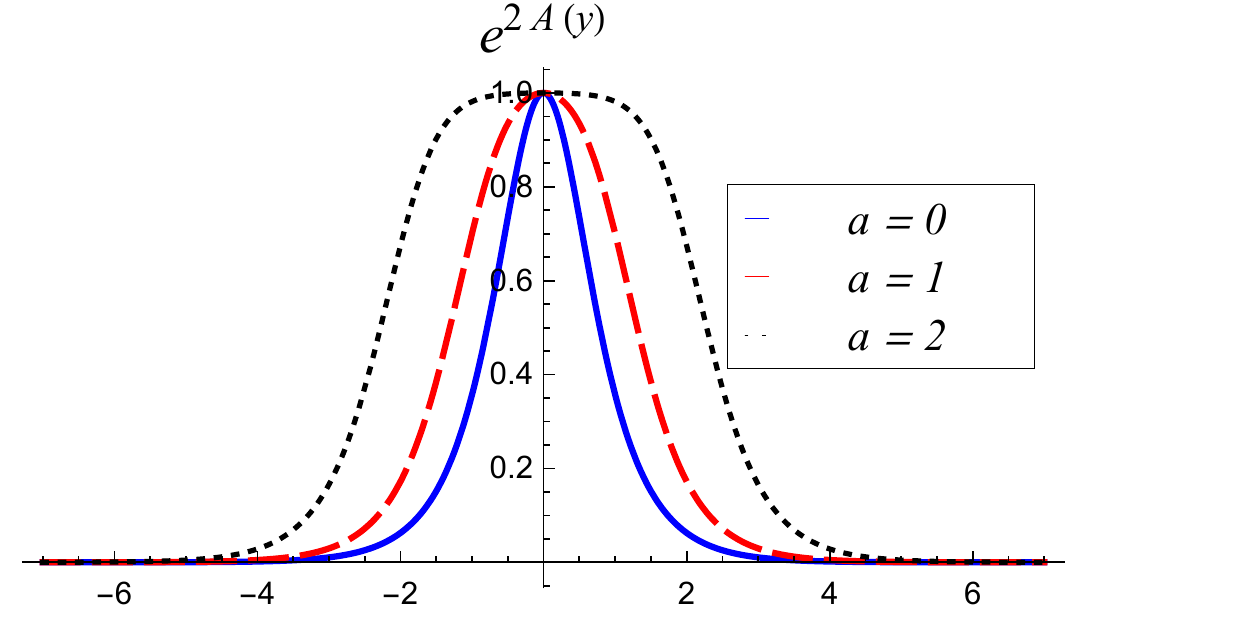}
        \includegraphics[scale=0.6]{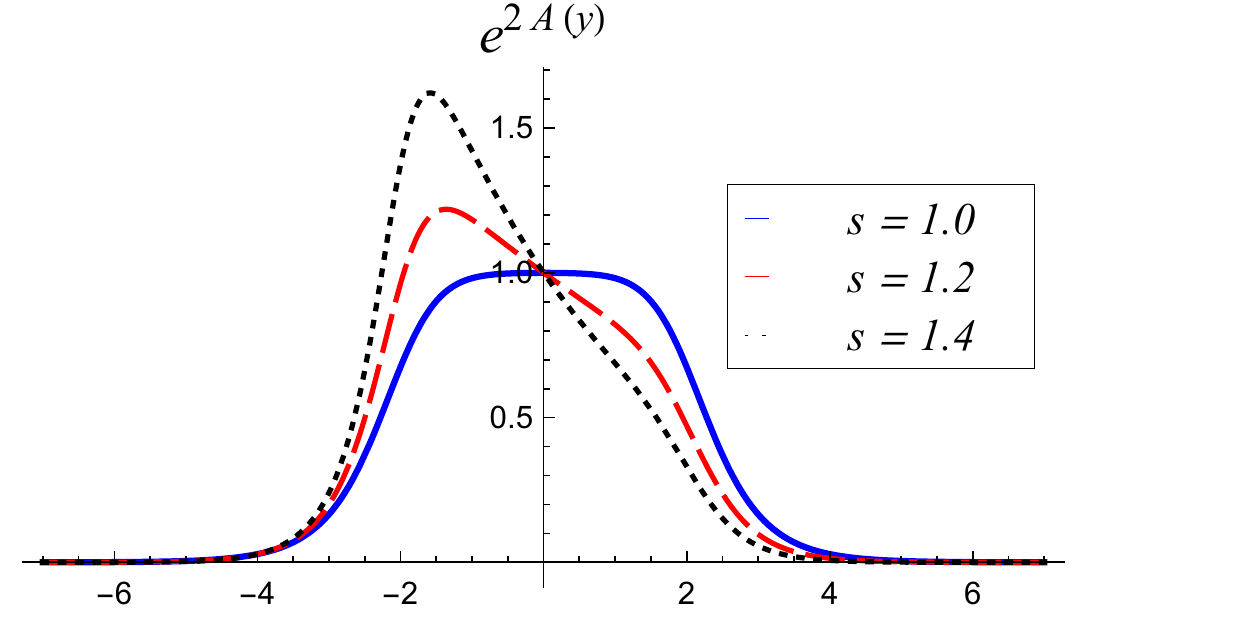}
    \end{center}
    \vspace{-0.5cm}
    \caption{\small{The upper panel show the warp factor for $r\!=\!s\!=\!1$, and for $a=0,1,$ and $2$. The lower panel show the warp factor for $r=1$, $a=2$ and for $s=1.0, 1.2,$ and $1.4$.}\label{fig1}}
\end{figure}

The solutions \eqref{soluMA} can be used to obtain the warp function as
\be\label{wfMA}
\begin{aligned}
A(y)=&\,\frac{4}{9}\ln\frac{\sech \big(r(y-a)\big)\,\sech \big(s(y+a)\big)}{\sech(ra)\,\sech (sa)}\\
&+\frac{1}{9} \sech^2\big(r(y-a)\big)-\frac{1}{9}\sech^2(ra)\\
&+\frac{1}{9} \sech^2\big(s(y+a)\big)-\frac{1}{9} \sech^2(sa)\,,
\end{aligned}
\ee
which obeys $A(0)=0$. In Fig. \ref{fig1}, we display the warp factor obtained from the above function and verify that if $a=0$, it has the usual bell shape profile. However, if $a\neq 0$ it engenders two interesting behaviors: the first is the enlargement of the bounded region below the curve as represented in the upper panel in Fig.~\ref{fig1}, where we used $r=s=1$ and $a=0,\,1$ and $2$; the second behavior is represented in the lower panel of Fig. \ref{fig1}, where we fixed $a=2$ and increased the thickness of the field $\chi$ by increasing the parameter $s$. We see in this case that the warp factor becomes asymmetric. Despite the significant changes in warp factor behavior, the Kretschmann scalar, obtained by $K(y)=40A'^4+16A''^2+32A'^2A''$, behaves adequately for the values of parameters which we used in this model.

\begin{figure}[t]
    \begin{center}
        \includegraphics[scale=0.6]{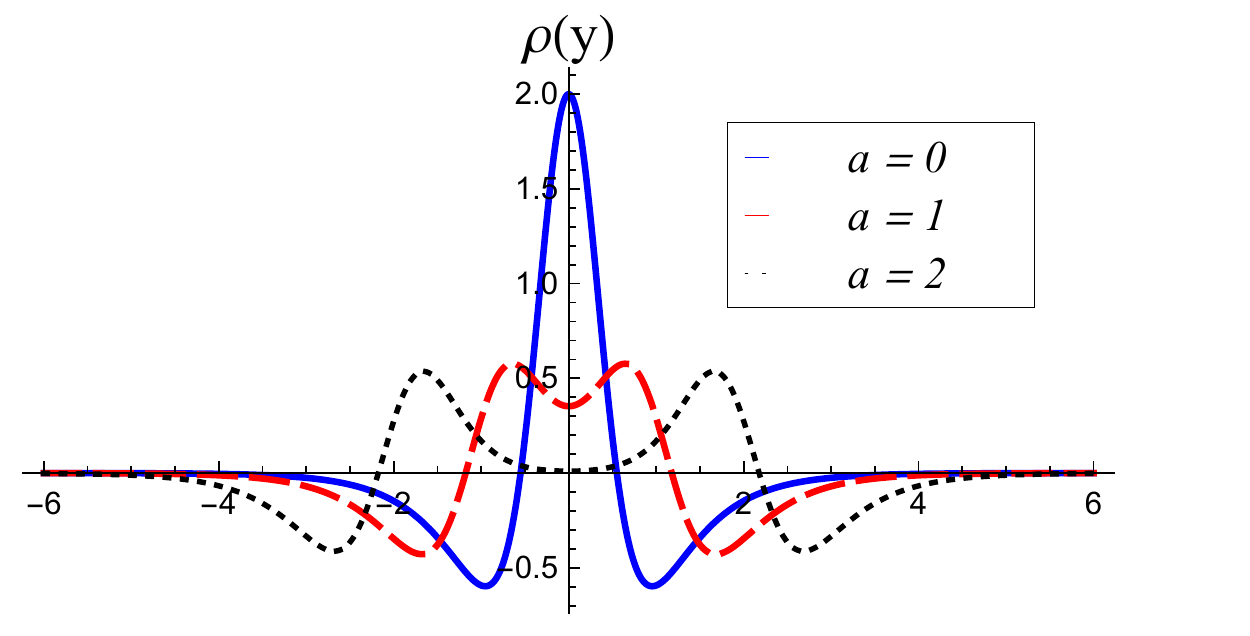}
        \includegraphics[scale=0.6]{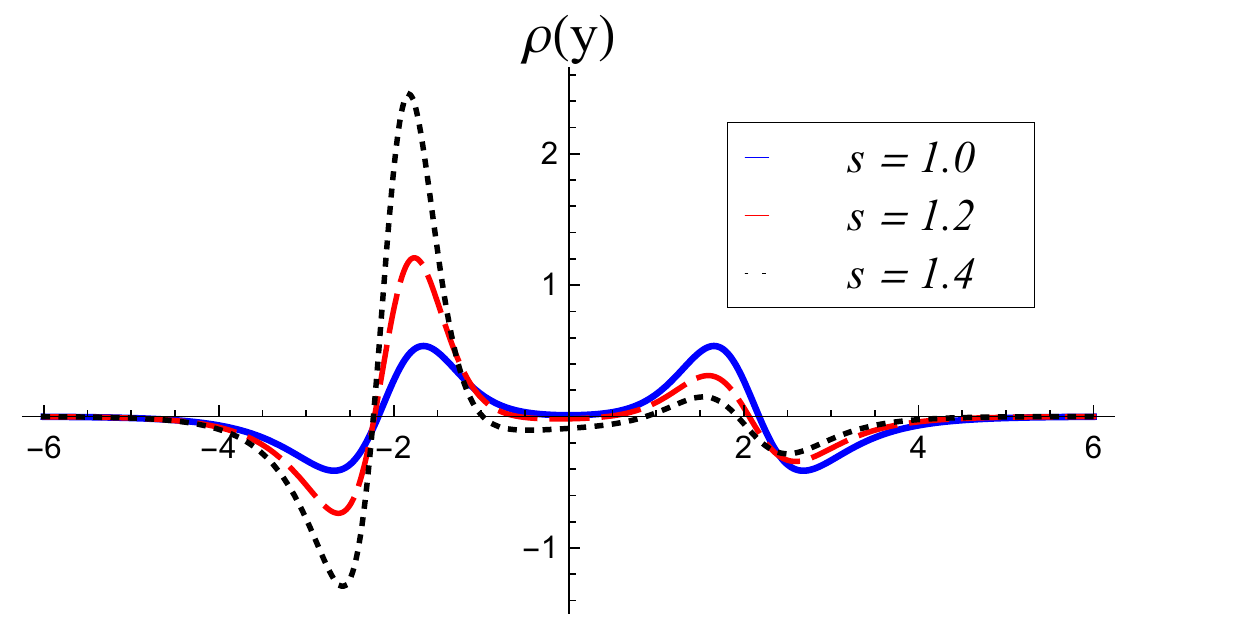}
    \end{center}
    \vspace{-0.5cm}
    \caption{\small{The upper panel show the energy density for $r\!=\!s\!=\!1$ and $a=0,1,$ and $2$. The lower panel show the energy density for $r=1$, $a=2$ and for $s=1.0, 1.2$ and $1.4$.}\label{fig2}}
\end{figure}

Using the Eq. \eqref{EnergyDensityFO} we can write the energy density in the form
\be\label{EnDenMA}
\begin{aligned}
\!\!\!\rho(y)=&\,r^2e^{2A} \sech^4(r(y\!-\!a)) \!+\! s^2e^{2A} \sech^4(s(y\!+\!a))\\
\!\!\!&-\frac{4e^{2A}}{3}\Big(r \tanh(r(y\!-\!a))-\frac{r}{3}\tanh^3(r(y\!-\!a))\\
\!\!\!&+s\tanh(s(y\!+\!a))-\frac{s}{3}\tanh^3(s(y\!+\!a))\!\Big)^2\,,
\end{aligned}
\ee
where $A(y)$ is given by Eq. \eqref{wfMA}. Fig. \ref{fig2} shows the energy density for the same values of the parameters used in Fig. \ref{fig1}. We can see in the upper panel of Fig. \ref{fig2} that the energy density goes from a local maximum at $y=0$ when $a=0$ to a minimum when $a\neq 0$, indicating that the distance between the centers of the solutions introduces an internal structure in the energy density. This result is similar to that obtained in the case of one-field models that engender two-kink solutions, as shown in \cite{Bazeia:2003aw}. However, in the present scenario we can increase the thickness of one of the solutions, which leads to an asymmetry in energy density. Despite these changes, the total energy of the brane remains zero.

\begin{figure}[t]
    \begin{center}
        \includegraphics[scale=0.6]{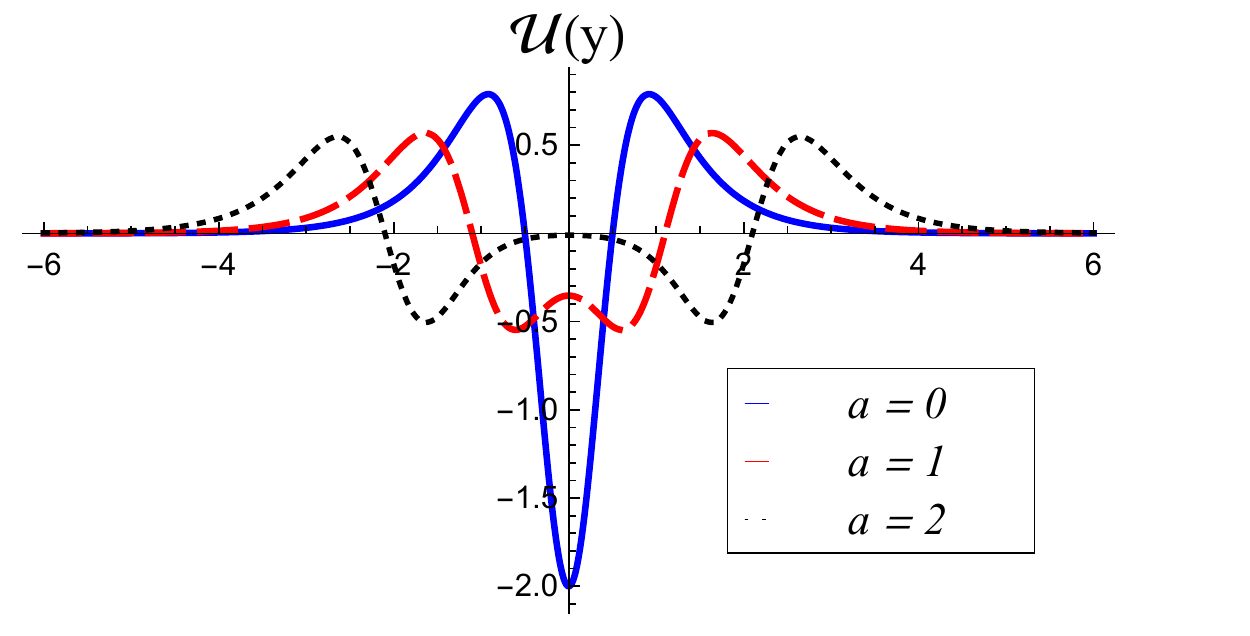}
        \includegraphics[scale=0.6]{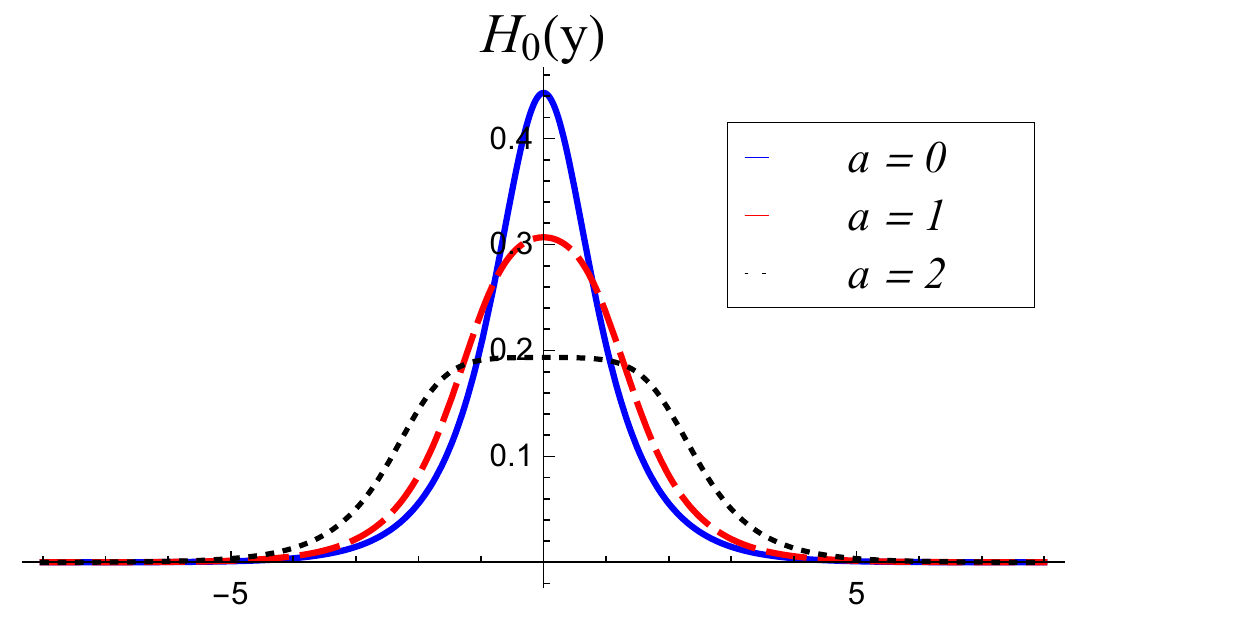}
    \end{center}
    \vspace{-0.5cm}
    \caption{\small{Stability potential (upper panel) and zero mode (lower panel) depicted for $r=s=1$ and $a=0,1$ and $2$.}\label{fig3}}
\end{figure}

\begin{figure}[t]
    \begin{center}
        \includegraphics[scale=0.6]{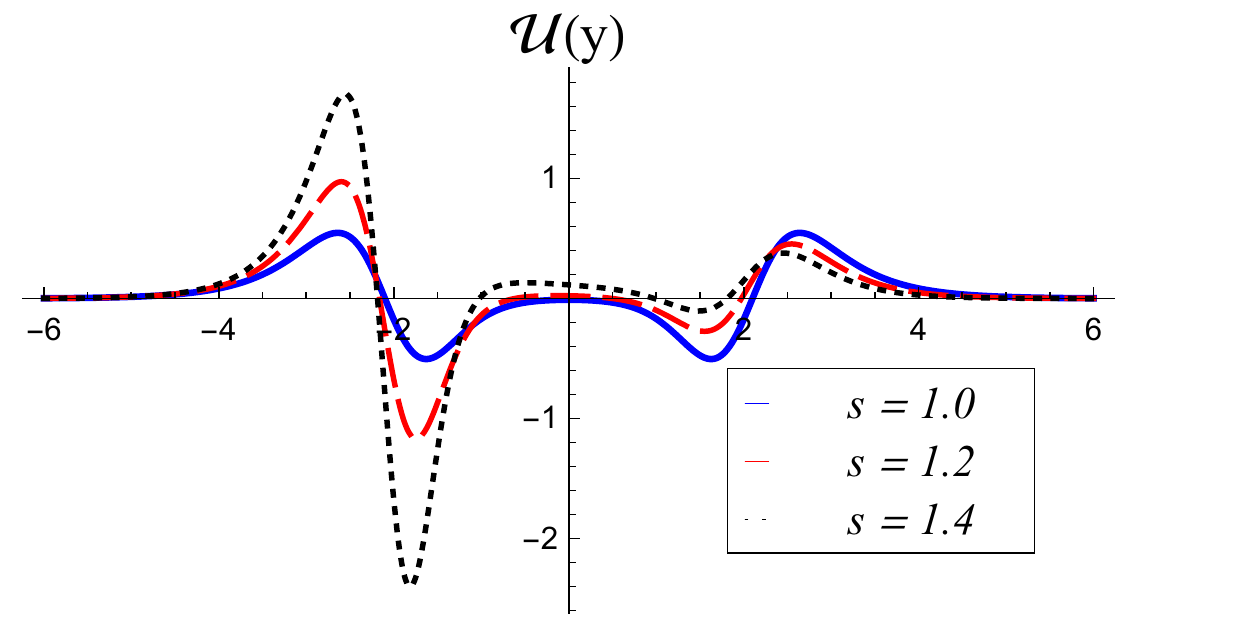}
        \includegraphics[scale=0.6]{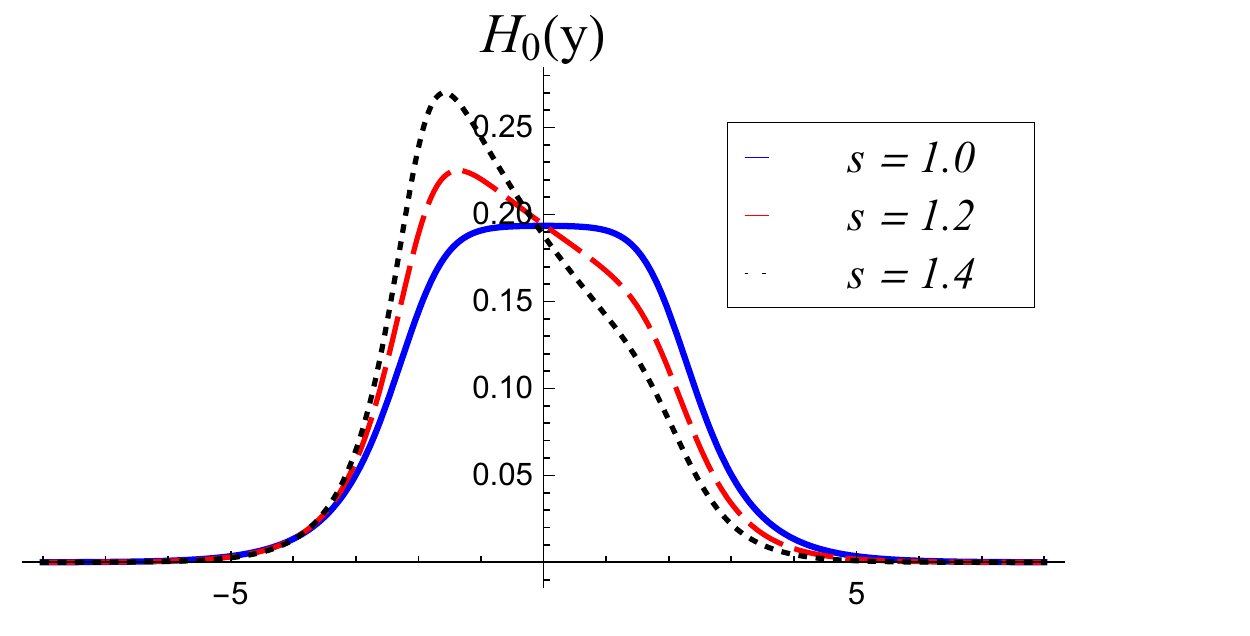}
    \end{center}
    \vspace{-0.5cm}
    \caption{\small{Stability potential (upper panel) and zero mode (lower panel) depicted for $r=1$ and $a=2$, and $s=1.0, 1.2$ and $1.4$.}\label{fig4}}
\end{figure}

We also verified the behavior of the stability potential and the zero mode. These results are also analytical, but we decided not to show them explicitly here because they are given by large and awkward expressions. However, we depicted them in Figs. \ref{fig3} and \ref{fig4}. As we can see in Fig. \ref{fig3}, the stability potential (upper panel) can have its central well divided in two, causing an enlargement on the localization of the zero mode (lower panel). Moreover, when one considers $a\neq 0$ and adds larger values for the parameter that controls one of the fields, there is an increase in the depth of one of the potential wells as shown in the upper panel of the Fig. \ref{fig4}. As a consequence, the zero mode tends to be more and more concentrated around the corresponding deeper well. As one can see, there are other possibilities, which will not be explored in the present work.

\begin{figure}[t]
    \begin{center}
        \includegraphics[scale=0.6]{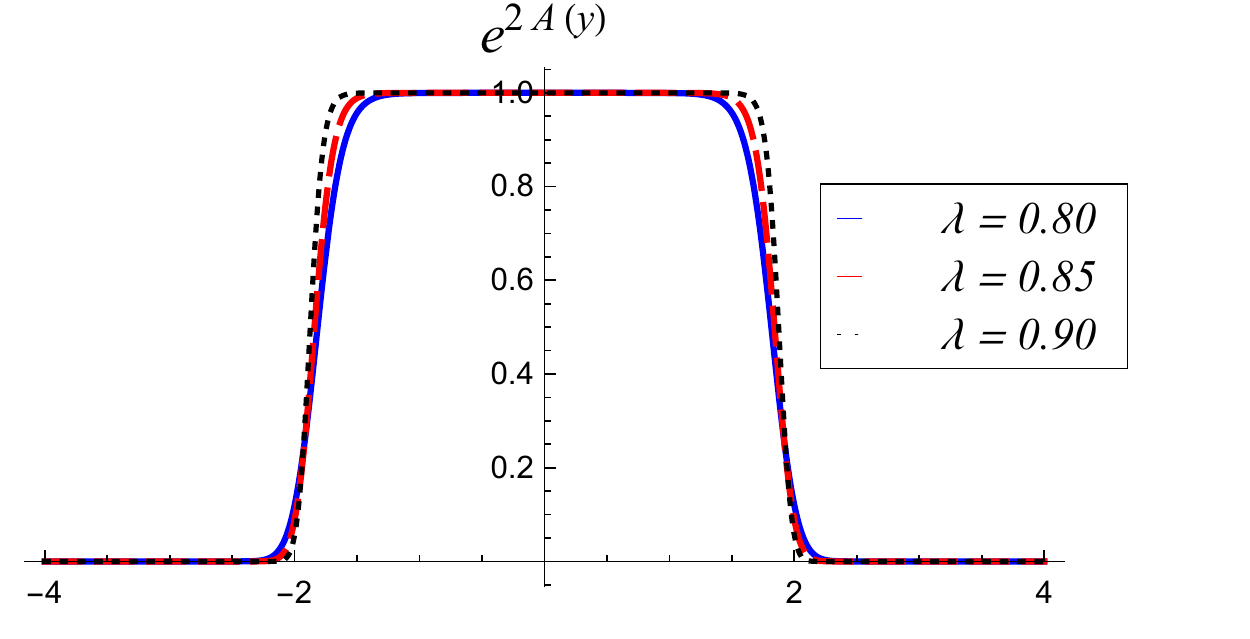}
        \includegraphics[scale=0.6]{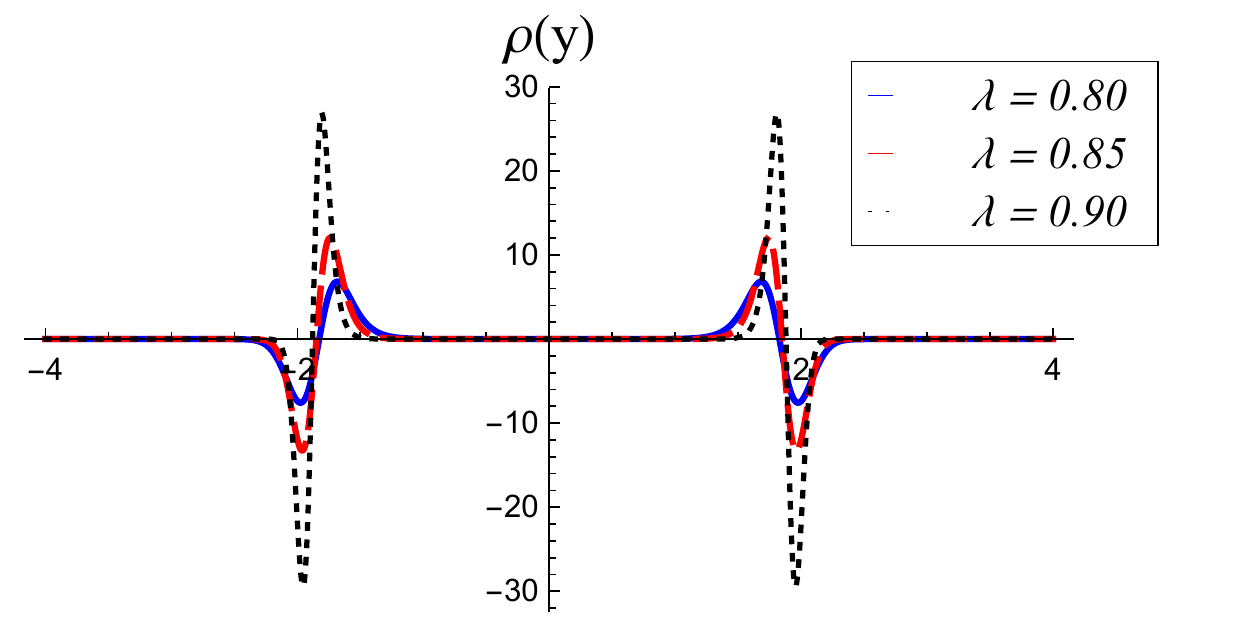}
    \end{center}
    \vspace{-0.5cm}
    \caption{\small{Warp factor (upper panel) and energy density (lower panel) depicted for $r=s=1$, $a=2$ and $\lambda=0.80, 0.85$ and $0.90$.}\label{figMD1}}
\end{figure}

Let us now investigate a new situation constructed by a two-field model that seems to induce a compactification of the geometry of the brane. For this, we follow Ref. \cite{Bazeia:2015eta} and consider
\ben
\begin{aligned}
\!\!\!W(\phi,\chi)=&\,-\frac{r}{\sqrt{\lambda}(1\!-\!\lambda)}\ln\!\left(\!\frac{1\!-\!\sqrt{\lambda}\,\sn(\phi,\lambda)}{\dn(\phi,\lambda)}\!\right)\\
\!\!\!&\,-\frac{s}{\sqrt{\lambda}(1\!-\!\lambda)}\ln\!\left(\!\frac{1\!-\!\sqrt{\lambda}\,\sn(\chi,\lambda)}{\dn(\chi,\lambda)}\!\right),
\end{aligned}
\een
where $\lambda$, $r$ and $s$ are real parameters, with $\lambda\in [0,1)$. Also, ${\sn}$ and $\dn$ are Jacobi's elliptic functions. Evidently, we could use distinct $\lambda_1$ and $\lambda_2$ to make the brane asymmetric, but here we consider the above model to check how the localization works in the new scenario; as shown in \cite{Bazeia:2015eta}, for $\lambda$ increasing towards unity, the solution shrinks to become more and more localized. In this new model, using the first-order equations we obtain the solutions as
\bes\label{soluMC}
\bal
\phi(y)&=\,\sn^{-1}\left(\tanh\bigg(\frac{r(y-a)}{1-\lambda}\bigg),\,\lambda\,\right),\\
\chi(y)&=\,\sn^{-1}\left(\tanh\bigg(\frac{s(y-b)}{1-\lambda}\bigg),\,\lambda\,\right).
\eal
\ees
Again, one takes $b=-a$ and in Fig \ref{figMD1} one shows the warp factor (upper panel) depicted for $r= s= 1$, $a= 2$ and $\lambda= 0.80, 0.85$ and $0.90$. Notice that as $\lambda$ increases, the localization becomes more and more accentuated. The lower panel in Fig. \ref{figMD1} show the energy density. Again, one notices that the energy density becomes more and more localized inside the compact interval $y\in[-2,2]$, as $\lambda$ increase towards unity.

In order to better understand the process of compactification, one can notice, for instance, from the behavior of the warp factor displayed in Fig. \ref{figMD1}, that the quantity
\be 
\Sigma_\lambda= 2\int_2^\infty dy\; e^{2A_\lambda(y)}
\ee 
which represents the area outside the compact interval $[-2,2]$, decreases as $\lambda$ increases towards unity. We have calculated this quantity for several values of $\lambda$, in particular, $\Sigma_{0.95}=6.70\cdot 10^{-4}$, $\Sigma_{0.97}=1.69\cdot 10^{-4}$, $\Sigma_{0.99}=6.63\cdot 10^{-6}$ and $\Sigma_{0.999}=2.10\cdot 10^{-9}$, evidencing the asymptotic compact behavior of the configuration.

\begin{figure}[h]
    \begin{center}
        \includegraphics[scale=0.6]{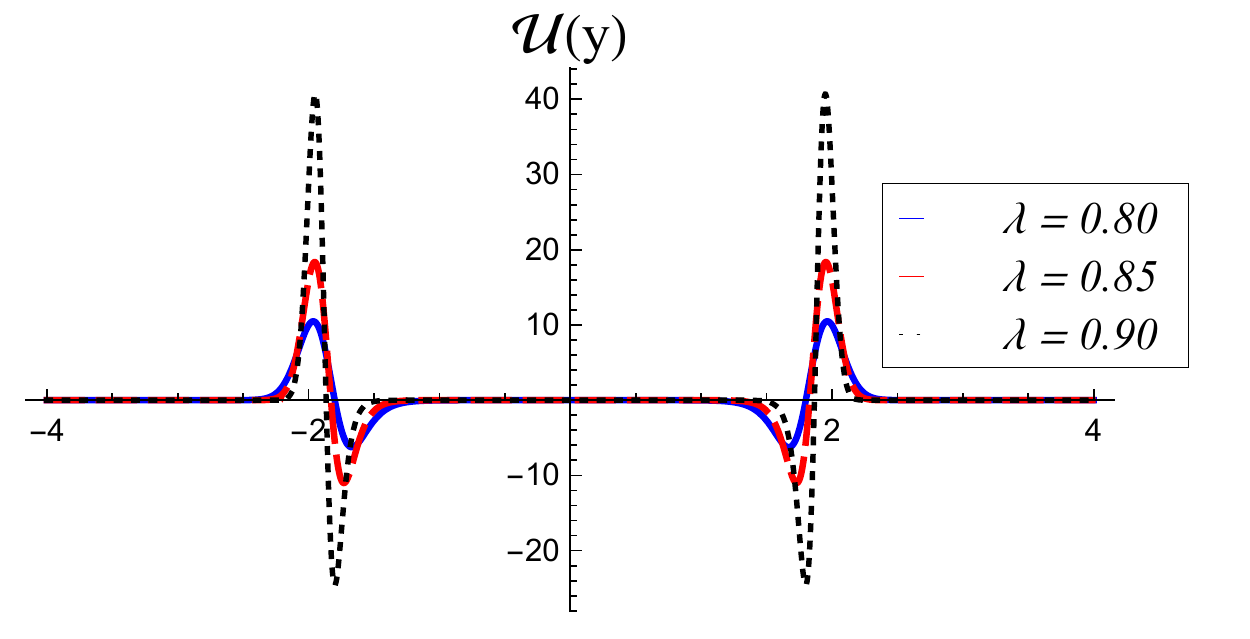}
        \includegraphics[scale=0.6]{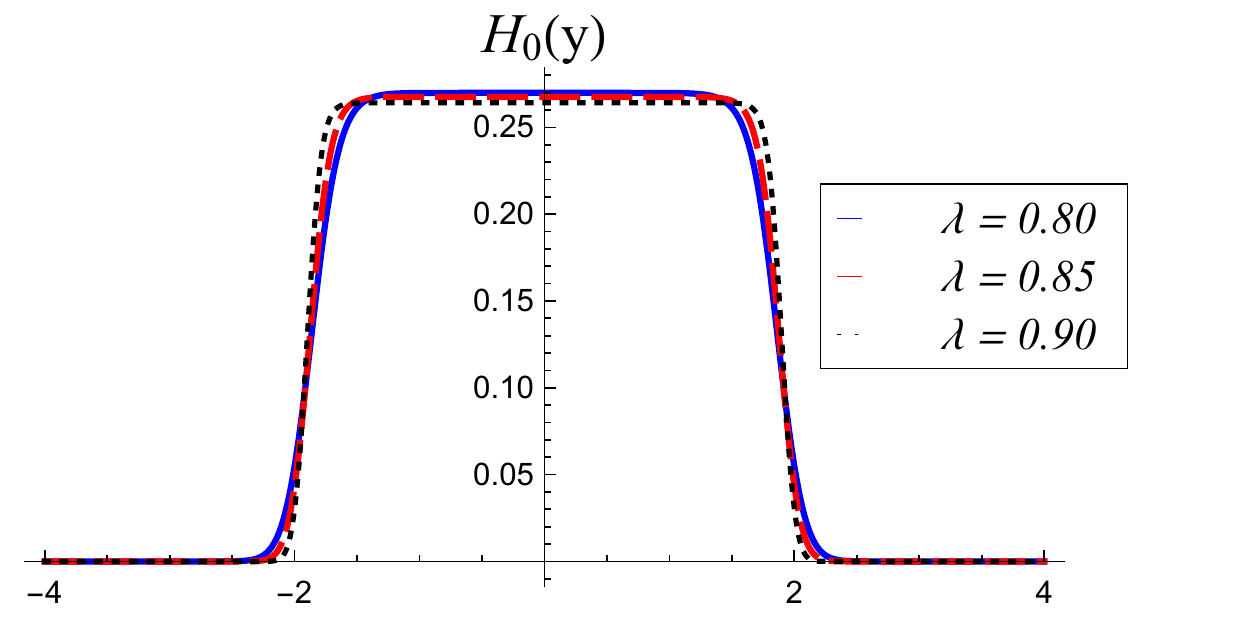}
    \end{center}
    \vspace{-0.5cm}
    \caption{\small{Stability potential (upper panel) and zero mode (lower panel) depicted for $r=s=1$, $a=2$ and $\lambda=0.80, 0.85$ and $0.90$.}\label{figMD3}}
\end{figure}

We also analyzed the behavior of the stability potential and zero mode in this new scenario and verified that compactification also changes their behavior as we can see in upper and lower panel of Fig. \ref{figMD3}.

\begin{center}{\it{Three fields}}\end{center}

\begin{figure}[t]
    \begin{center}
        \includegraphics[scale=0.6]{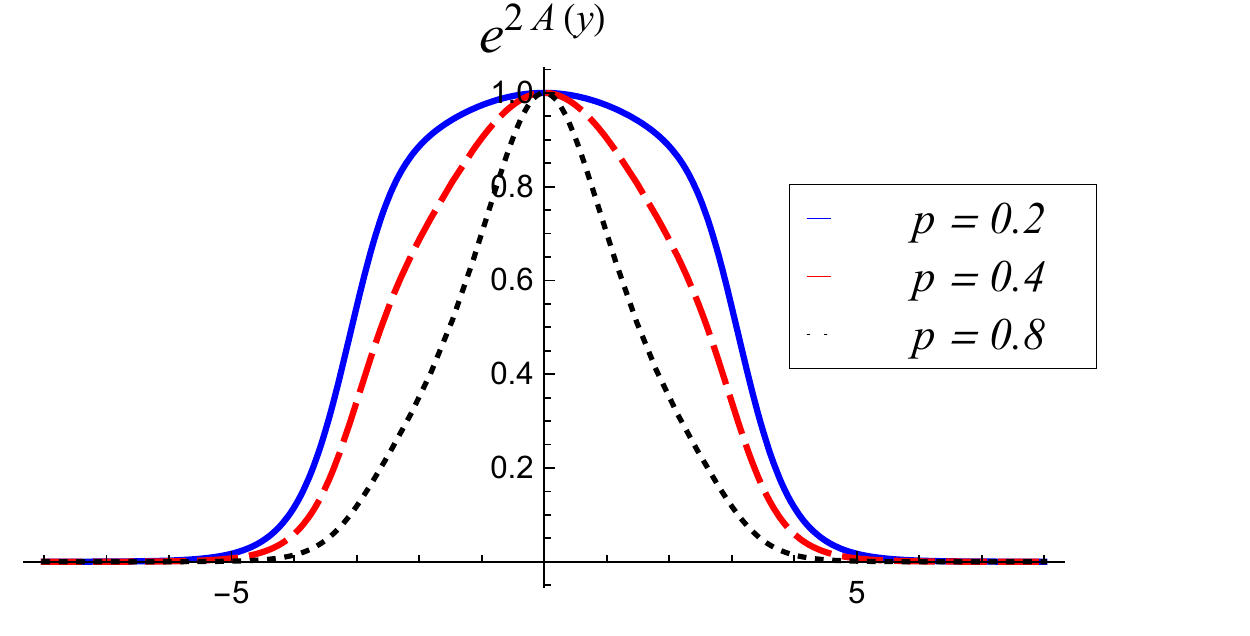}
        \includegraphics[scale=0.6]{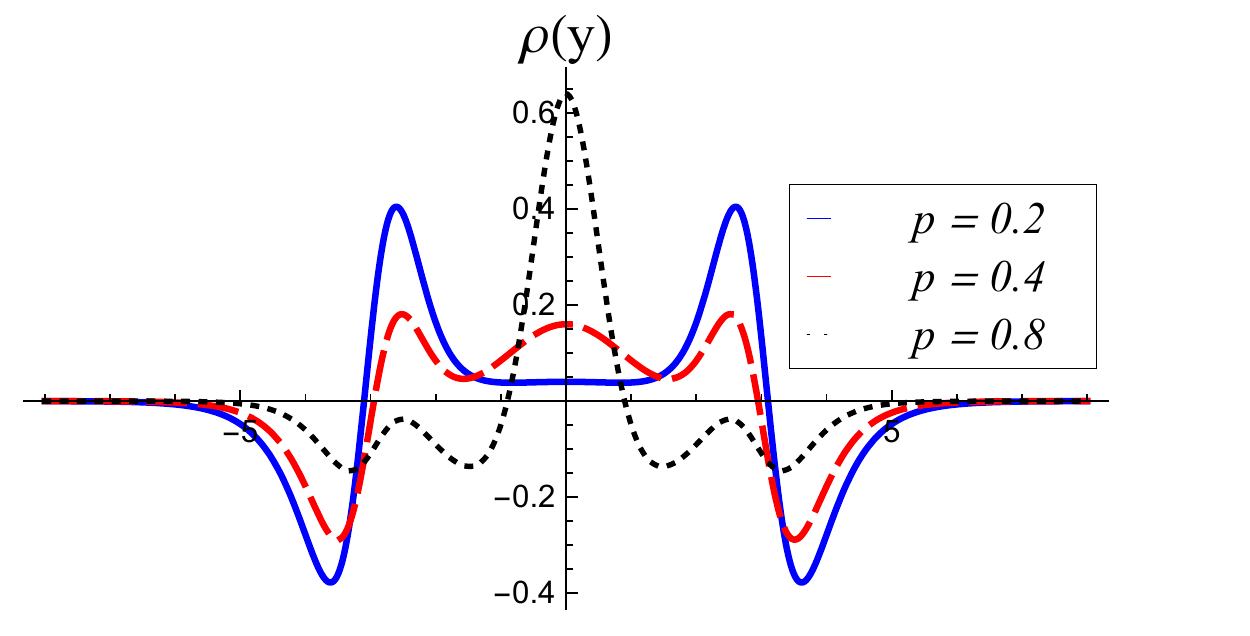}
    \end{center}
    \vspace{-0.5cm}
    \caption{\small{Warp factor (upper panel) and energy density (lower panel) depicted for $r=s=1$, $a=3$ and $p=0.2, 0.4$ and $0.8$.}\label{fig5}}
\end{figure}

Let us now investigate another model, containing the three real scalar fields $\phi$, $\chi$ and $\psi$. In this case, we consider the auxiliary function $W$ in the form
\be
W(\phi,\chi,\psi)\!=\!r\!\left(\!\phi\!-\!\frac13\phi^3\!\right)\!+s\!\left(\!\chi\!-\!\frac13\chi^3\!\right)\!+p\!\left(\!\psi\!-\!\frac13\psi^3\!\right)\!,
\ee 
where the new parameter $p$ controls the field $\psi$. The potential $V$ is given by
\be
\begin{aligned}
V(\phi,\chi,\psi)&=\,\frac{r^2}2\!\left(1\!-\!\phi^2\right)^2\!+\!\frac{s^2}2\!\left(1\!-\!\chi^2\right)^2\!+\!\frac{p^2}2\!\left(1\!-\!\psi^2\right)^2\\
&-\frac43\!\left(\!r\!\left(\!\phi\!-\!\frac13\phi^3\!\right)\!+\!s\!\left(\!\chi\!-\!\frac13\chi^3\!\right)\!+\!p\!\left(\!\psi\!-\!\frac13\psi^3\!\right)\!\right)^{\!2}\!\!,
\end{aligned}
\ee
and the first-order equations are
\bes\label{EPOMB}
\bal
\phi'&=r\left(1-\phi^2\right),\\
\chi'&=s\left(1-\chi^2\right),\\
\psi'&=p\left(1-\psi^2\right).
\eal
\ees
The solutions can be written as
\bes\label{soluMB}
\bal
\phi(y)&=\,\tanh\left(r(y-a)\right),\\
\chi(y)&=\,\tanh\left(s(y-b)\right),\\
\psi(y)&=\,\tanh\left(p(y-c)\right),
\eal
\ees
where $a$, $b$ and $c$ are real parameters. Here we take $b=-a$ and $c=0$, to centralize the field $\psi$ at $y=0$. As we have checked, the set of solutions \eqref{soluMB} also solve the second-order equations \eqref{fieldeqs}. Moreover, the warp function has the form
\be\label{wfMB}
\begin{aligned}
\!\!\!A(y)=&\,\frac{4}{9}\ln\frac{\sech \big(r(y-a)\big)\,\sech \big(s(y+a)\big)\,\sech(py)}{\sech(ra)\,\sech (sa)}\\
\!\!\!&+\frac{1}{9} \sech^2\big(r(y-a)\big)+\frac{1}{9} \sech^2\big(s(y+a)\big)\\
\!\!\!&-\frac{1}{9} \tanh^2(p y)-\frac{1}{9}\sech^2(ra)-\frac{1}{9} \sech^2(sa)\,,
\end{aligned}
\ee
which obeys $A(0)=0$. In the upper panel in Fig. \ref{fig5}, we depict the warp factor for $r=s=1$, $a=3$ and $p=0.2,0.4$ and $0.8$. We can also get an asymmetric warp factor, in this case increasing the thickness of the fields $\phi$ and $\chi$, controlled by $r$ and $s$.

The energy density has the form
\be\label{EnDenMB}
\begin{aligned}
\!\!\!\rho(y)=&\,r^2e^{2A} \sech^4(r(y\!-\!a)) \!+\! s^2e^{2A} \sech^4(s(y\!+\!a))\\
\!\!\!&+p^2e^{2A} \sech^4(py) -\frac{4e^{2A}}{3}\Big(r\tanh(r(y\!-\!a))\\
\!\!\!&-\frac{r}{3}\!\tanh^3(r(y\!-\!a))+s\tanh(s(y\!+\!a))\\
\!\!\!&-\frac{s}{3}\!\tanh^3(s(y\!+\!a))\!+\!p \tanh(py)\!-\frac{p}{3}\!\tanh^3(py)\!\Big)^2.
\end{aligned}
\ee
It is displayed in Fig. \ref{fig5} and one sees that changes in the thickness of the third field significantly modifies the core structure of energy density, indicating modification of the internal structure of the model.

\begin{figure}[t]
    \begin{center}
        \includegraphics[scale=0.6]{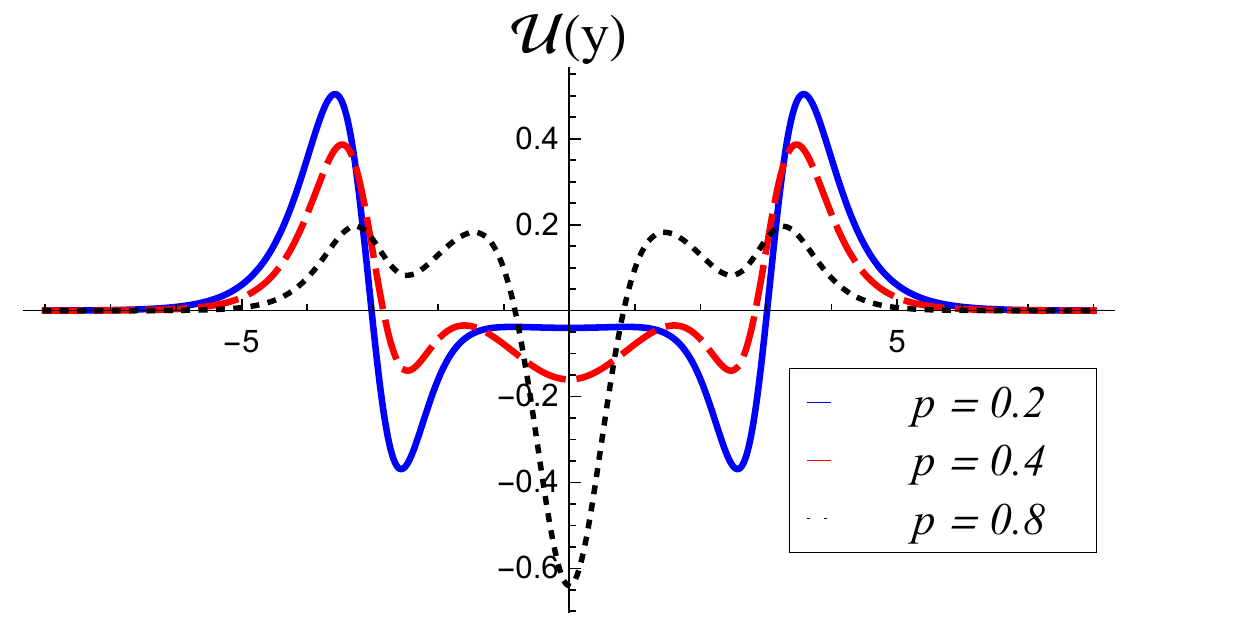}
        \includegraphics[scale=0.6]{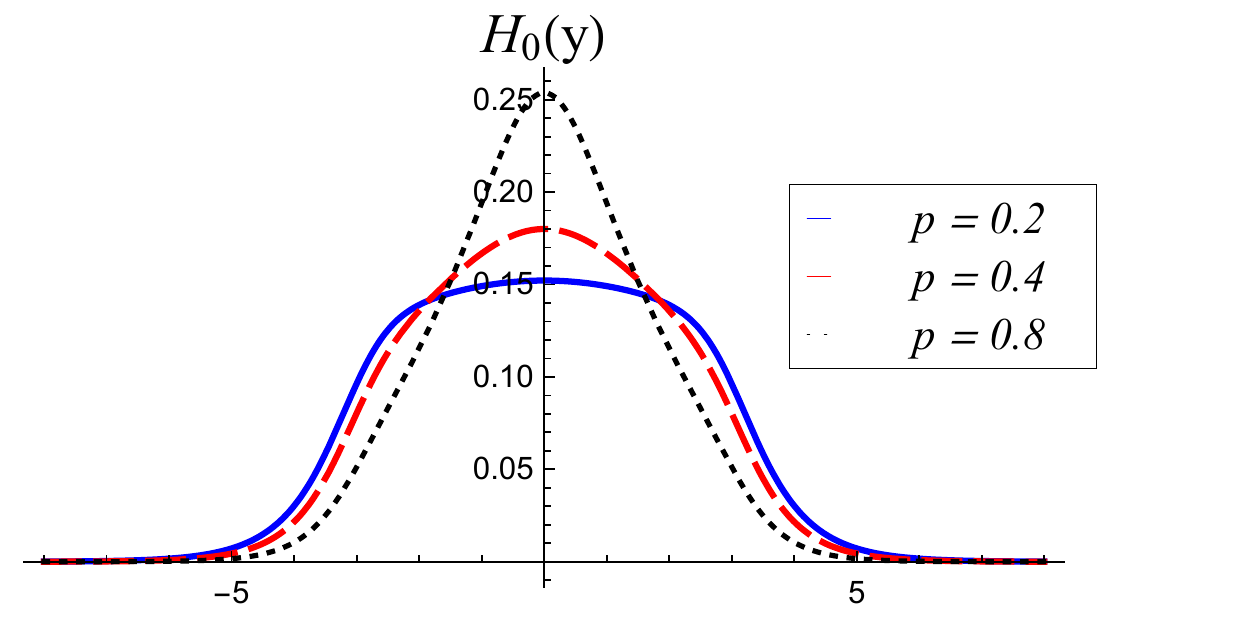}
    \end{center}
    \vspace{-0.5cm}
    \caption{\small{Stability potential (upper panel) and zero mode (lower panel) depicted for $r=s=1$, $a=3$ and $p=0.2, 0.4$ and $0.8$.}\label{fig6}}
\end{figure}

We also verify the stability potential and the zero mode in this case. The solutions are all analytic, and in Fig. \ref{fig6} we depict the results for $r=s=1$, $a=3$ and $p=0.2,0.4,$  and $0.8$, as in Fig. \ref{fig5}. Note that the stability potential can have a global minimum surrounded by two local minima (dotted-line), or three degenerate minima (dashed-line) or it can have two symmetric minima shifted from the center (solid-line). This makes the zero mode more or less localized around the center of the brane.

We further illustrate the use of three scalar fields investigating another model. This new possibility explores the presence of kink-like configurations that are strongly located around its centers, as  originally studied in \cite{Bazeia:2015eta}. We considered $W$ as
\ben
\begin{aligned}
\!\!\!W(\phi,\chi,\psi)=&\,-\frac1{\sqrt{\lambda}(1\!-\!\lambda)}\ln\!\left(\!\frac{1\!-\!\sqrt{\lambda}\,\sn(\phi,\lambda)}{\dn(\phi,\lambda)}\!\right)\\
\!\!\!&\,-\frac1{\sqrt{\lambda}(1\!-\!\lambda)}\ln\!\left(\!\frac{1\!-\!\sqrt{\lambda}\,\sn(\chi,\lambda)}{\dn(\chi,\lambda)}\!\right)\\
\!\!\!&\,+r\left(\psi+\frac13\psi^3\right),
\end{aligned}
\een
where $\lambda$ and $r$ are real parameters, with $\lambda\in [0,1)$. Also, ${\sn}$ and $\dn$ are Jacobi's elliptic functions.

\begin{figure}[t]
    \begin{center}
        \includegraphics[scale=0.6]{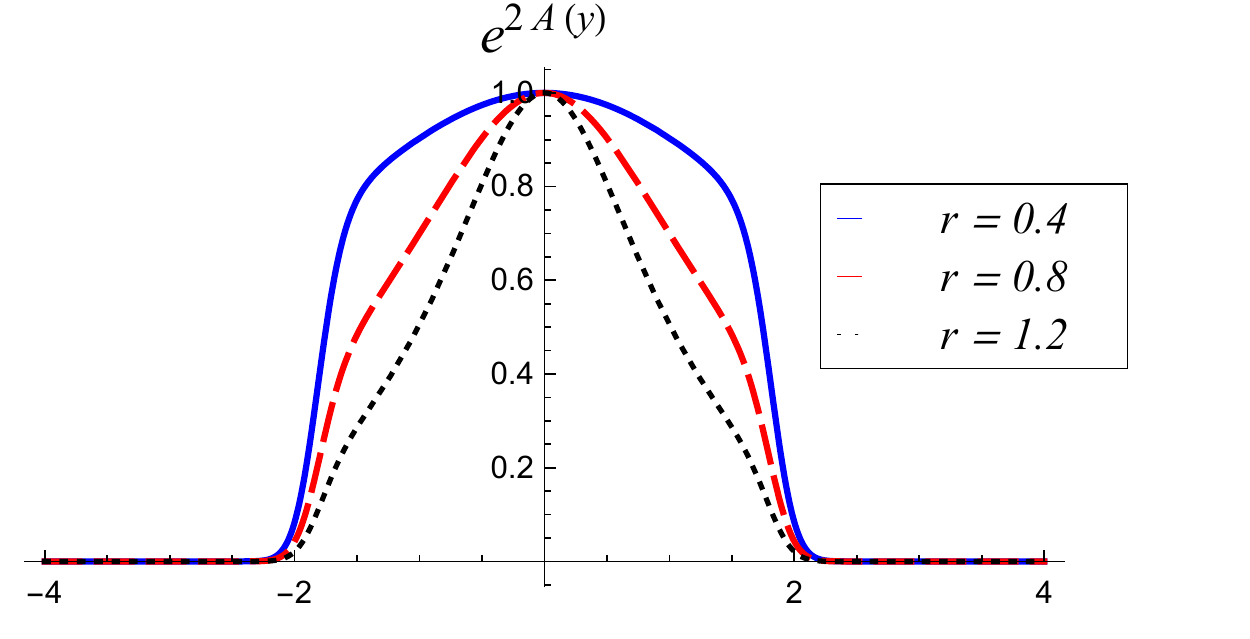}
        \includegraphics[scale=0.6]{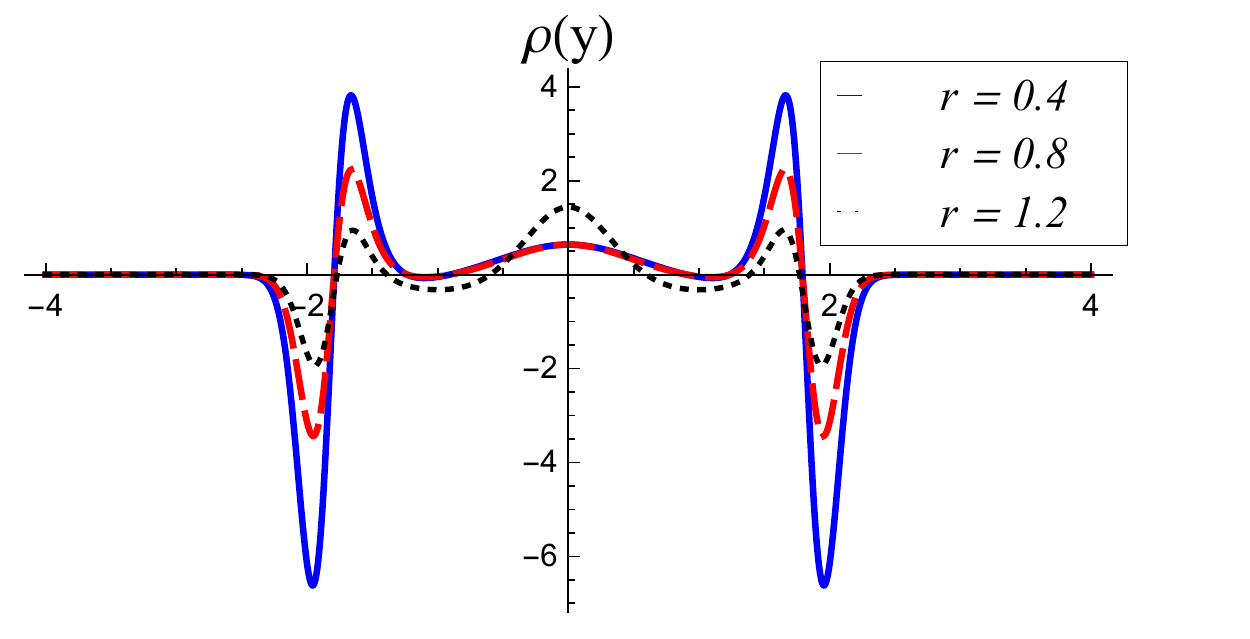}
    \end{center}
    \vspace{-0.5cm}
    \caption{\small{Warp factor (upper panel) and energy density (lower panel) depicted for $\lambda=0.8$, $a=2$ and $r=0.4, 0.8$ and $1.2$.}\label{fig7}}
\end{figure}

The solutions of the scalar fields are obtained as
\bes\label{soluMC}
\bal
\phi(y)&=\,\sn^{-1}\left(\tanh\bigg(\frac{y-a}{1-\lambda}\bigg),\,\lambda\,\right),\\
\chi(y)&=\,\sn^{-1}\left(\tanh\bigg(\frac{y-b}{1-\lambda}\bigg),\,\lambda\,\right),\\
\psi(y)&=\,\tanh\left(r(y-c)\,\right).
\eal
\ees
Again, one takes $b=-a$ and $c=0$. The model is more involved and the warp factor is obtained numerically. It is shown in Fig. \ref{fig7}, together with the corresponding energy density, using $a=2$. As we see, both quantities are strongly influenced by the positions of the two lateral kink-like configurations, located at $y=\pm2$.

\begin{figure}[h]
    \begin{center}
        \includegraphics[scale=0.6]{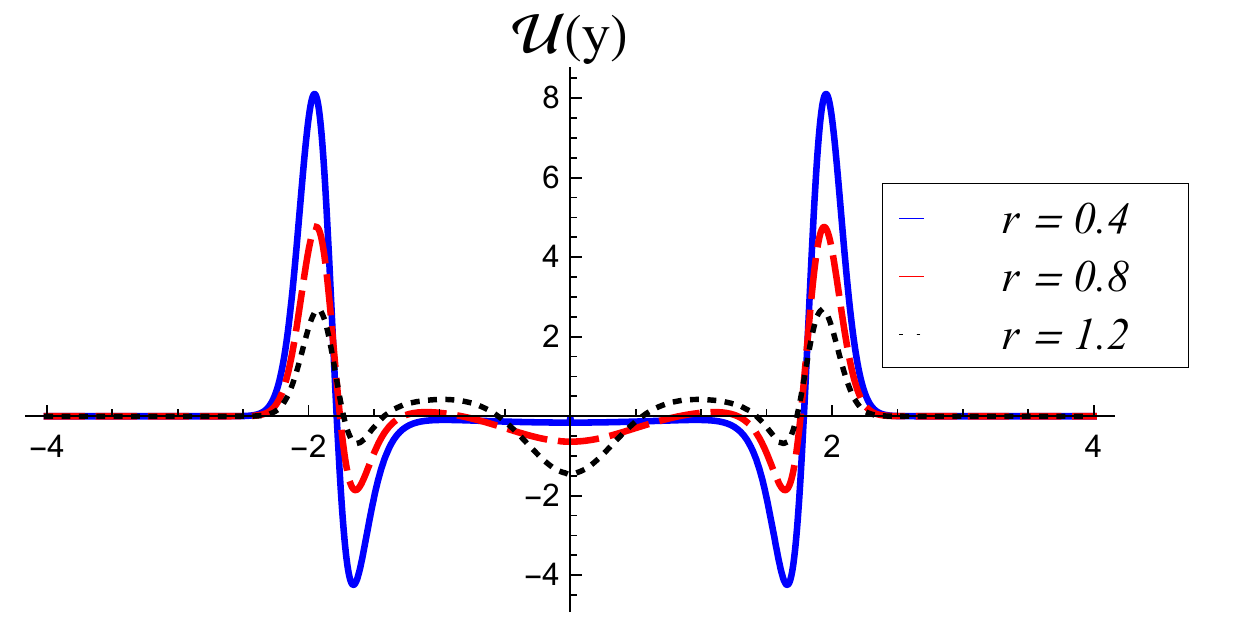}
        \includegraphics[scale=0.6]{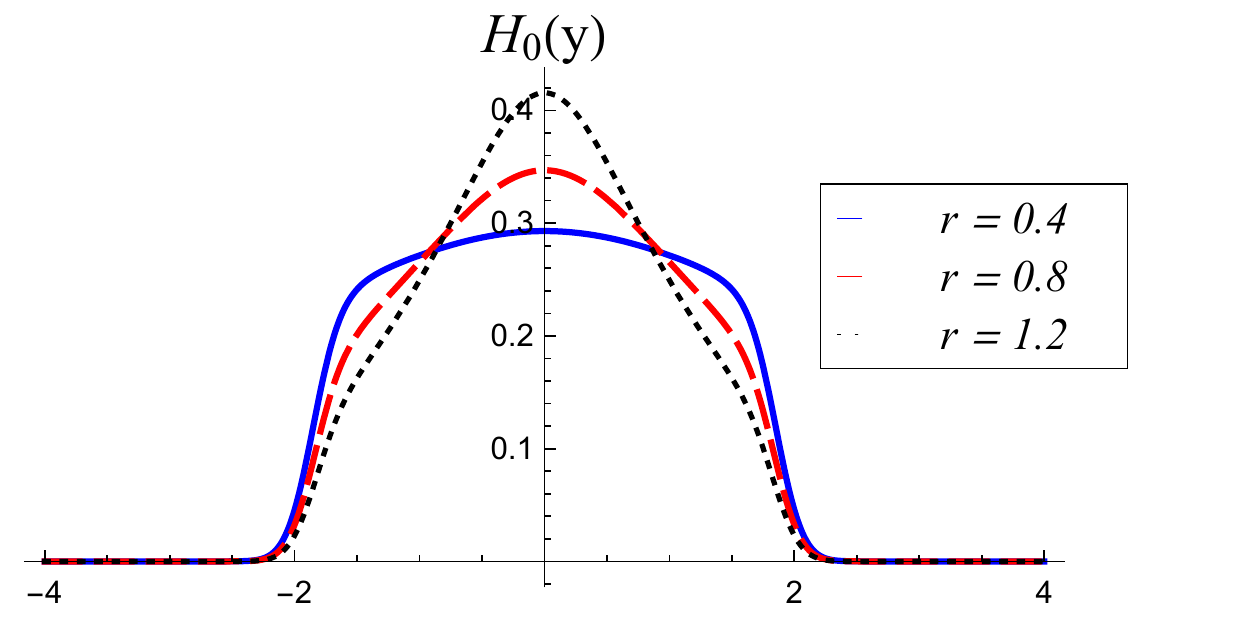}
    \end{center}
    \vspace{-0.5cm}
    \caption{\small{Stability potential (upper panel) and zero mode (lower panel) depicted for $\lambda=0.8$, $a=2$ and $r=0.4, 0.8$ and $1.2$.}\label{fig8}}
\end{figure}

The stability potential and zero mode were also studied, and they are displayed in the upper and lower panel of Fig. \ref{fig8}, respectively. As one can see, it is possible to construct several other possibilities, which will be explored elsewhere.

{\it{4. Conclusion.}}\, In this work we studied braneworld models sourced by several real scalar fields. We studied four specific models where the fields are described by kink-like solutions. In each situation we assumed that the centers of the solutions were shifted symmetrically by a parameter $a$, causing the solutions to interact in an unusual way. We analyzed the effects of the displacement of the center of the solutions on the warp factor, energy density, stability potential and the associated zero mode. In the first two-field model, the solutions obtained are parameterized to control the thickness and location of the brane. We noticed that when the kink-like solutions are located at the same point, the brane profile is similar to that obtained in one-field models. However, when one chooses solutions that are shifted from each other, the brane may become thicker, in a way similar to models with double-kink solutions \cite{Bazeia:2003aw}. But in the present context, we can also modify the action of one of the fields by increasing its thickness, which creates an asymmetry in the brane, inducing a novel behavior. Specifically, when we shifted the center of the kink-like configurations, also increasing the thickness of one of the fields, the stability potential acquired an asymmetric double well profile, with the zero mode tending to become increasingly concentrated in the potential well that is most affected by the increase in the thickness of one of the solutions. In the second two-field model, we investigated the possibility to shrink the brane inside a compact interval. These effects are also new, and we have not seen them before.

In the first three-field model, we analyzed the effects of the parameters on the center of the kink-like solutions. For simplicity, we chose to locate the center of the solution of one of the fields at $y=0$, shifting the center of the others two fields symmetrically. Here, new representations for the energy density and stability potential appeared. Firstly, the energy density acquired novel internal structure, allowing a transition between a minimum or a maximum at $y=0$, surrounded by two maxima shifted from the origin. Moreover, the stability potential engendered a central well surrounded by two others, or only two wells displaced from the center, leading to a zero mode which may be more or less localized around the origin. These results are also new, involving a novel mechanism that directly controls the internal profile of the brane. 

In the second three-field model, we introduced another potential, using kink-like configurations that can be strongly located at specific positions, also contributing to strongly change the internal structure of the brane.

The results unveiled a novel mechanism to control the braneworld, significantly changing the behavior of important quantities such as the warp factor, energy density, stability potential and zero mode. The parameters that adjust the center of the kink-like solutions directly contribute to change the internal structure of the model, leading to interesting new profiles. A direct issue that deserves further examination concerns the confinement of fermion and gauge fields inside the brane, and how they can contribute phenomenologically; see, e.g., Refs. \cite{Correa:2010zg,Zhao:2011hg,Cruz:2011ru,Zhang:2016ksq} and references therein. Another issue is related to the presence of asymmetry and cosmic acceleration, which can be considered as in \cite{Padilla:2004tp,Brito:2005zw}. The modifications in the structure of the brane and in the gravitational stability induce new behavior, and this can also be studied in the context of modified gravity such as $F(R)$ and $F(R,T)$. In the scalar-tensor representation of $F(R,T)$, we can follow the lines of Refs. \cite{Rosa:2020uli,Rosa:2021tei}, for instance, and also \cite{Rosa:2022fhl} to include compact or asymmetric information. Beyond the braneworld context, we can turn attention to cosmology, to study the extension of the first-order formalism developed in Ref. \cite{cosmo} to the case of several scalar fields in a way similar to the one studied in the present work. These and other related issue are presently under investigation, and we hope to report on them in the near future.\\

\noindent{\bf Acknowledgements:} The authors would like to thank Conselho Nacional de Desenvolvimento Cient\'\i fico e Tecnol\'ogico, CNPq, and Paraiba State Research Foundation, FAPESQ-PB, grant No. 0015/2019, for partial financial support. DB also thank CNPq, grant No. 303469/2019-6, for financial support.

\noindent{{\bf Declaration of competing interest:} The authors declare that they have no known competing financial interests or personal relationships that could have
appeared to influence the work reported in this paper.}

\noindent{\bf Data Availability Statement:} This manuscript has no associated data
or the data will not be deposited. [Authors’ comment: This theoretical
work does not use or produce numerical data.]


\end{document}